\documentclass[aps,prl,twocolumn,superscriptaddress,a4paper]{revtex4}

\usepackage{graphicx}

\pdfoutput=1 %Needed for ArXiv submission

\usepackage{xfrac}
\usepackage{fancyhdr}
\usepackage[colorlinks=true,citecolor=blue,linkcolor=magenta]{hyperref}
\hypersetup{%
pdfauthor = {Patrick Maletinsky},
colorlinks = true, linkcolor = blue, urlcolor=blue, bookmarksnumbered =  true}

\usepackage{xfrac}

\newcommand{\ket}[1]{\left|#1\right>}

\begin{document}

% Use the \preprint command to place your local institutional report
% number in the upper righthand corner of the title page in preprint mode.
% Multiple \preprint commands are allowed.
% Use the 'preprintnumbers' class option to override journal defaults
% to display numbers if necessary
%\preprint{}

%Title of paper
\title{Nanoscale magnetic imaging of a single electron spin under ambient conditions}

% repeat the \author .. \affiliation  etc. as needed
% \email, \thanks, \homepage, \altaffiliation all apply to the current
% author. Explanatory text should go in the []'s, actual e-mail
% address or url should go in the {}'s for \email and \homepage.
% Please use the appropriate macro for each each type of information

% \affiliation command applies to all authors since the last
% \affiliation command. The \affiliation command should follow the
% other information
% \affiliation can be followed by \email, \homepage, \thanks as well.
\author{M.S. Grinolds\footnote{These authors contributed equally to this work}}
\affiliation{Department of Physics, Harvard University, Cambridge, Massachusetts 02138 USA}
\author{S. Hong\footnotemark[\value{footnote}]}
\affiliation{School of Engineering and Applied Science, Harvard University, Cambridge, Massachusetts, 02138 USA}
%\email[]{grinolds@fas.harvard.edu}
\author{P. Maletinsky\footnotemark[\value{footnote}]}
\affiliation{Department of Physics, University of Basel, Klingelbergstrasse 82, Basel CH-4056, Switzerland}
%\email[]{patrickm@physics.harvard.edu}
\author{L. Luan}
\affiliation{Department of Physics, Harvard University, Cambridge, Massachusetts 02138 USA}
\author{M.D. Lukin}
\affiliation{Department of Physics, Harvard University, Cambridge, Massachusetts 02138 USA}
\author{R.L. Walsworth}
\affiliation{Department of Physics, Harvard University, Cambridge, Massachusetts 02138 USA}
\affiliation{Harvard-Smithsonian Center for Astrophysics, Cambridge, Massachusetts 02138 USA}
\author{A. Yacoby}
\affiliation{Department of Physics, Harvard University, Cambridge, Massachusetts 02138 USA}
\email[]{yacoby@physics.harvard.edu}
%\homepage[]{Your web page}
%\thanks{}

\date{\today}

\begin{abstract}
The detection of ensembles of spins under ambient conditions has revolutionized the biological, chemical, and physical sciences through magnetic resonance imaging\,\cite{Mansfield2004} and nuclear magnetic resonance\,\cite{Rabi1938,Bloch1946}. Pushing sensing capabilities to the individual-spin level would enable unprecedented applications such as single molecule structural imaging; however, the weak magnetic fields from single spins are undetectable by conventional far-field resonance techniques4. In recent years, there has been a considerable effort to develop nanoscale scanning magnetometers\,\cite{Martin1987,Chang1992,Zuger1993,Kirtley1995}, which are able to measure fewer spins by bringing the sensor in close proximity to its target. The most sensitive of these magnetometers generally require low temperatures for operation, but measuring under ambient conditions (standard temperature and pressure) is critical for many imaging applications, particularly in biological systems. Here we demonstrate detection and nanoscale imaging of the magnetic field from a single electron spin under ambient conditions using a scanning nitrogen-vacancy (NV) magnetometer. Real-space, quantitative magnetic-field images are obtained by deterministically scanning our NV magnetometer $50~$nanometers above a target electron spin, while measuring the local magnetic field using dynamically decoupled magnetometry protocols. This single-spin detection capability could enable single-spin magnetic resonance imaging of electron spins on the nano- and atomic scales and opens the door for unique applications such as mechanical quantum state transfer.
\end{abstract}

\maketitle

To date, the magnetic fields from single electron spins have only been imaged under extreme conditions (ultralow temperatures and high vacuum), with data integration times on the order of days\,\cite{Rugar2004}. Magnetometers based on negatively charged nitrogen-vacancy (NV) centers in diamond have been proposed as sensors capable of measuring individual spins\,\cite{Balasubramanian2008,Degen2008,Maze2008} \,\cite{Taylor2008} because they can be initialized and read-out optically\,\cite{Gruber1997} and have long coherence times\,\cite{Balasubramanian2009}, even under ambient conditions. Moreover, since NV centers are atomic in size, they offer significant advantages in magnetic resolution and sensing capabilities if they can be brought in close proximity of targets to be measured. Recent advances in diamond nanofabrication have allowed for the creation of robust scanning probes that host individual NV centers within roughly $25~$nm of their tips\,\cite{Maletinsky2012}. Here, we employ such a scanning NV center to image the magnetic dipole field of a single target electron spin. 

\begin{figure}
\includegraphics[scale=1]{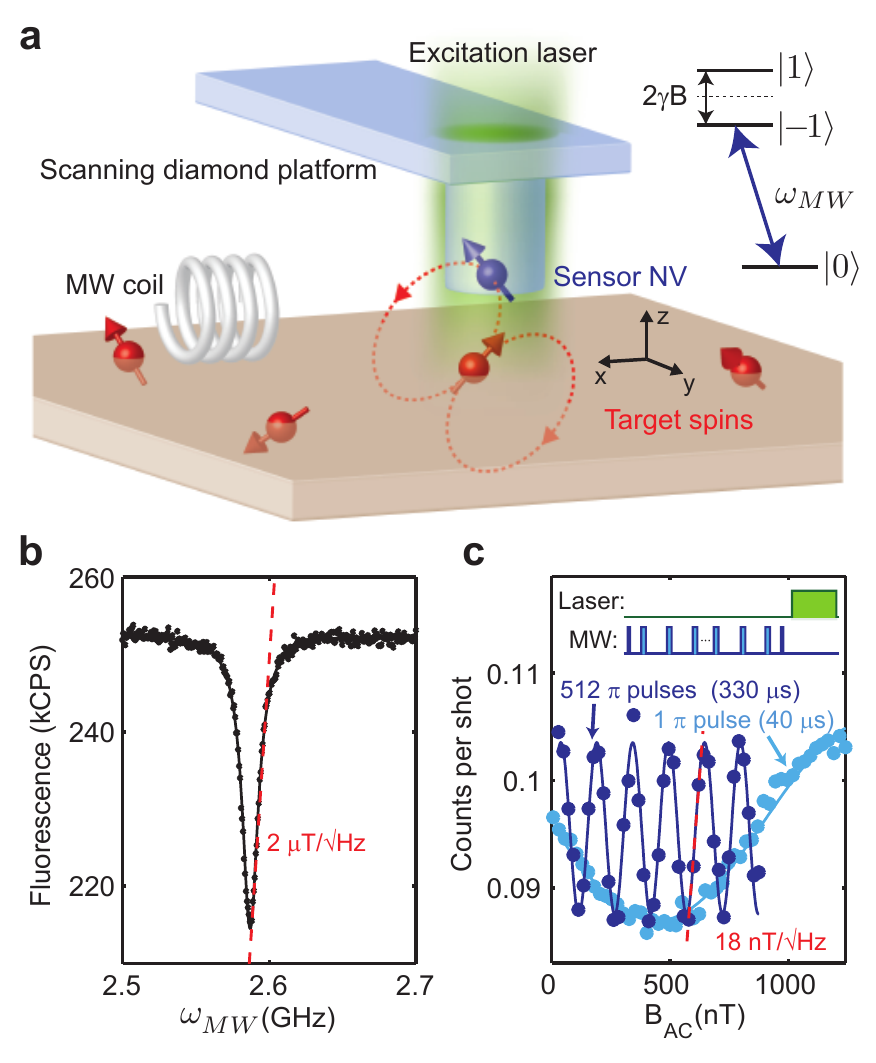}
\caption{\label{Fig1} Scanning NV magnetometer. (a) Conceptual schematic of the scanning NV magnetometer. The sensor NV is hosted within a scanning diamond nanopillar\,\cite{Maletinsky2012}, where its spin is initialized and read-out optically from above ($532~$nm excitation laser spot shown). Coherent NV spin manipulations are performed via a nearby microwave (MW) coil, in this work operating near resonance with the $\ket{0}$ or $\ket{-1}$ transition, in the presence of a static applied magnetic field (not shown). The sensor NV is scanned over target spins of interest to construct magnetic field images. (b) By continuously applying the excitation and sweeping the MWs across the $\ket{0}$ to $\ket{-1}$ transition, optically detected magnetic resonance provides a measure of the static magnetic field at the NV center, with a DC sensitivity of $\sim2\mu$T/$\sqrt{\rm Hz}$. (c) By dynamically decoupling the sensor NV spin from its environment, the sensor's magnetic field sensitivity is dramatically improved for AC magnetic fields. Plotted are a spin-echo (1-pulse) magnetometry sequence with a $40-\mu$s total evolution time, and a $512$-pulse XY$8$\,\cite{Gullion1991} magnetometry sequence (see Supplementary Discussion) with $330-\mu$s total evolution time, which achieve magnetic field sensitivities of $56~$nT/$\sqrt{\rm Hz}$ and 18 nT/$\sqrt{\rm Hz}$, respectively.}
\end{figure}

	Our scanning NV magnetometer (Fig.\,\ref{Fig1}a) consists of a combined confocal and atomic force microscope (AFM) as previously described\,\cite{Grinolds2011}, which hosts a sensing NV center embedded in a diamond nanopillar scanning probe tip\,\cite{Maletinsky2012}. The sensor NV's spin-state is initialized optically and read out through spin-dependent fluorescence, while its position relative to the sample is controlled through atomic-force feedback between the tip and sample. Microwaves (MWs) are used to coherently manipulate the sensor NV spin. Magnetic sensing is achieved by measuring the NV spin's optically detected electron spin resonance (ESR), either by continuously applying near-resonant MW radiation (Fig.\,\ref{Fig1}b) or through pulsed spin-manipulation schemes\,\cite{Maze2008,Taylor2008}, (Fig.\,\ref{Fig1}c), where the sensor NV spin precesses under the influence of its local magnetic field (projected along the NV center's crystallographic orientation). We measure the contribution of the magnetic field from a target electron spin to this precession. The entire system, including both the scanning NV magnetometer and the target sample, operates under ambient conditions. 

To verify the single-spin detection and imaging, we choose our target to be the spin associated with an additional negatively charged NV center in a separate diamond crystal (so that the sensor and target NV centers can be scanned relative to one another). The advantage of using an NV target is that both its location and spin state can be independently determined by its optical fluorescence. As discussed below, we can thus compare the target NV's magnetically measured location to its optically measured location and ensure that the magnetic image is from a single targeted spin. Additionally, we can guarantee that the target spin is initialized and properly modulated, as is useful for optimizing AC magnetic sensing.
	
To isolate single NV targets for imaging, NV centers are created in a shallow ($<25~$nm) layer of a bulk diamond through established implantation and annealing techniques, as used in previous work\,\cite{Grinolds2011}. The target diamond surface is structured to create nanoscale mesas, whose diameters ($\sim200~$nm) are chosen to contain, on average, a single NV spin. Mesas with single NV centers (as determined through photon auto-correlation experiments, Supplementary Fig.\,\ref{FigS1}) are chosen for our measurements. In order to individually control the target and sensor NV spins, we choose a target NV center with a different crystallographic orientation (which determines the spin quantization axis) from the sensor NV, so that their spin transitions can be spectrally separated in ESR measurements by applying a uniform static magnetic field.
	
\begin{figure}
\includegraphics[scale=1]{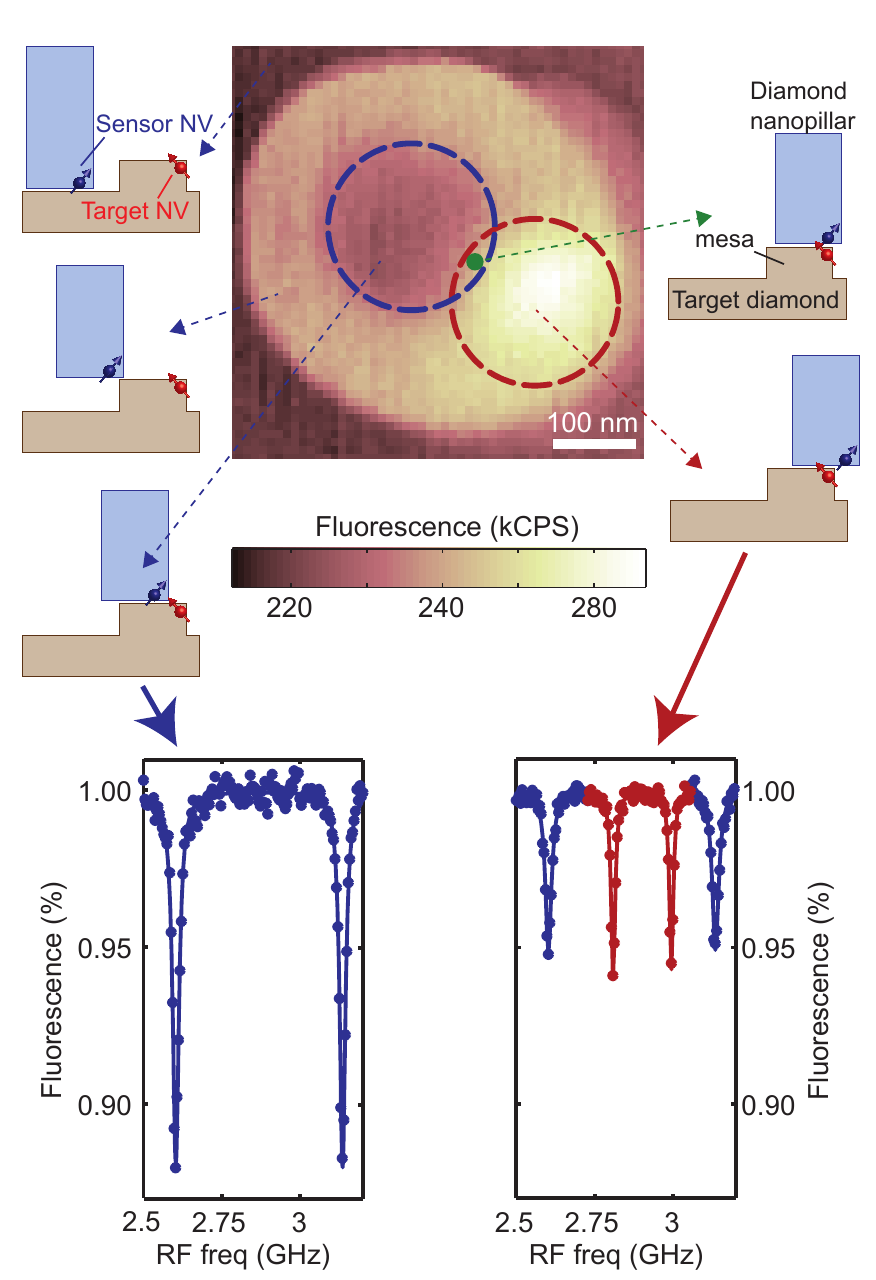}
\caption{\label{Fig2} Independent determination of a target spin's location. The sensor NV�s diamond nanopillar is scanned over a target nanostructure ("mesa") containing a single target NV center. The combined NV fluorescence is recorded as a function of position (top, center panel). The fluorescence has a strong spatial dependence because (i) sensor NV fluorescence can partially couple into the target bulk diamond when the sensor NV is close to the sample surface, and (ii) target NV fluorescence can couple into the nanopillar waveguide when the nanopillar is located above of the target NV. When the nanopillar is located away from the target NV, only fluorescence from the sensor NV is collected, as indicated by ESR measurements showing two spectral peaks corresponding to the sensor NV spin m$_S=0\leftrightarrow\pm1$ transitions (bottom, left panel). For ESR measurements taken with the nanopillar located above the target NV (bottom, right panel) there are four observable spectral peaks that correspond to both the sensor and target NV spin transitions (blue and red, respectively), with reduced ESR contrast due to collecting fluorescence from both NV spins. The center of the target-coupling circle (red dashed circle around bright fluorescence spot) indicates the lateral location of the target NV spin relative to the center of the nanopillar. Similarly, the center of the sensor-quenching circle (blue dashed circle around dark fluorescence spot) indicates the lateral location of the sensor NV spin. With both NV spins� lateral locations known, the position of sensor-target closest approach can be ascertained (green dot).}
\end{figure}

Spatial features in the collected fluorescence from scanning the NV magnetometer over target diamond mesas allow us to independently determine the relative positions of the sensor and target NV spins (Fig.\,\ref{Fig2}). Firstly, the scanning diamond nanopillar acts as a waveguide\,\cite{Babinec2010} which, when centered precisely above the target NV, provides efficient collection of fluorescence from the target NV (in addition to the sensor NV). Also, the sensor NV's fluorescence is efficiently coupled into the target bulk diamond when it is centered on a mesa, due to the diamond�s high refractive index. The intersection of these two near-field fluorescence features indicates where the sensor NV spin is closest to the target NV spin. This spatial location is later used to confirm the location of the magnetically imaged target NV spin.

\begin{figure*}
\includegraphics[scale=1]{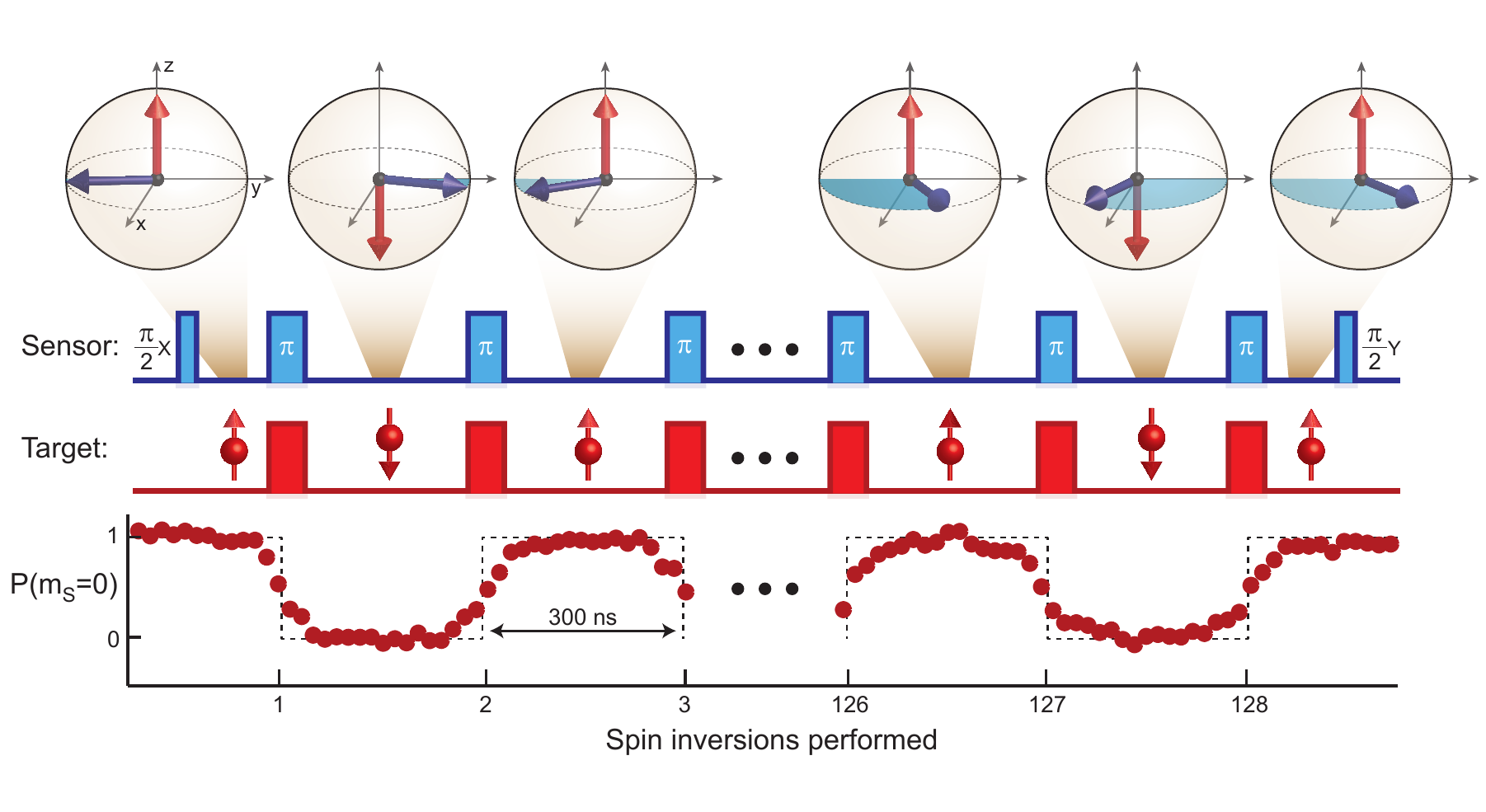}
\caption{\label{Fig3} Single spin detection scheme and target spin modulation verification. To detect the magnetic field from the target NV spin, the sensor NV spin (top panel, blue arrow) is placed in a superposition of spin states with a MW $\sfrac{\pi}{2}$ pulse (around the X axis). It then evolves under the influence of the magnetic field from the target spin (red arrow), accumulating phase (shaded blue region, whose extent is visually exaggerated for visibility). To optimize the sensor spin�s magnetic sensitivity, it is dynamically decoupled from its environment (see Supplementary Discussion)\,\cite{DeLange2010} by the repeated application of MW $\pi$-pulses using an XY$8$ sequence\,\cite{Gullion1991}. In order to magnetically measure the target NV spin, it is inverted, synchronously with the $\pi$-pulses applied to the sensor NV, so that phase shifts induced on the sensor by the target spin constructively accumulate. (The target NV spin is modulated between the m$_S=0$ and m$_S=-1$ states to isolate an effective spin-$\sfrac{1}{2}$ system from the target NV's spin triplet.) To invert the target NV spin with high fidelity, we employ adiabatic fast passages (see Supplementary Discussion). Plotted is the measured fluorescence for pulses $1$, $2$, $127$, and $128$, indicating that the target NV spin can be modulated many times without substantial polarization decay. The sensor NV's accumulated phase is converted to a population difference using a final $\sfrac{\pi}{2}$ pulse, whose axis (Y) is chosen to maximize sensitivity to small magnetic fields.}
\end{figure*}

Near the expected location of the target, the local magnetic field is measured with a magnetometry pulse sequence performed on the sensor NV using a combination of dynamic decoupling\,\cite{DeLange2010} and double electron-electron resonance\,\cite{Larsen1993}. The sensor NV spin is prepared in a superposition of spin states, where it accumulates phase proportional to the local magnetic field, including contributions from the target NV spin. To optimize magnetic field sensitivity, the sensor NV is dynamically decoupled from fluctuating magnetic fields in its environment (Fig.\,\ref{Fig3}, upper panel) through the repeated application of MW $\pi$-pulses. Normally, this pulse sequence would also remove any magnetic signal from a static target spin, but we also simultaneously invert the target NV spin in phase with the $\pi$-pulses applied to the sensor NV spin (Fig.\,\ref{Fig3}, lower panel) to maintain the sign of phase accumulation by the sensor spin due to the target NV spin. The total acquired phase is converted to a population difference, which is measured via NV spin-dependent fluorescence. 
	
With the sensor NV in close proximity to the target diamond surface, the field sensitivity of the sensor NV is reduced, because sensor NV fluorescence is partially emitted into the target bulk diamond (due to its high refractive index), and the target NV adds background fluorescence to magnetic measurements. Because of these effects, our sensor NV's magnetic field sensitivity at closest approach to the target NV is somewhat reduced to approximately $96~$nT/$\sqrt{\rm Hz}$ (with a $32$-$\pi$-pulse XY$8$\,\cite{Gullion1991} decoupling scheme and a $40$-$\mu$s total phase accumulation time; Supplementary Fig.\,\ref{FigS5}). Since the target NV is embedded in bulk diamond, the sensor-to-target vertical separation is roughly twice the distance between the sensor NV and the diamond surface. Thus, for our magnetic field imaging of a single target NV spin, we expect a $\sim50$-nm sensor-target vertical separation, which results in a magnetic field of about $10~$nT at the sensor NV location.
	
A magnetic field image centered at the expected target spin location is acquired by averaging the sensor's NV fluorescence in multiple scans of the NV magnetometer across a $\sim200\times200$-nm field-of-view (taken using a lateral drift correction scheme detailed in Supplementary Fig.\,\ref{FigS3}). A normalization scheme is applied to the magnetometry, where we alternately initialize the target NV spin in the $\ket{0}$ state and the $\ket{-1}$ state and measure the equal and opposite phase shifts induced during the sensor NV's magnetometry sequence (Supplementary Fig.\,\ref{FigS2}). We subtract the measured NV fluorescence rates for these two initial target NV spin polarizations, which isolates the magnetic field signal from the target spin (Supplementary Fig.\,\ref{FigS4}).
	
\begin{figure}
\includegraphics[scale=1]{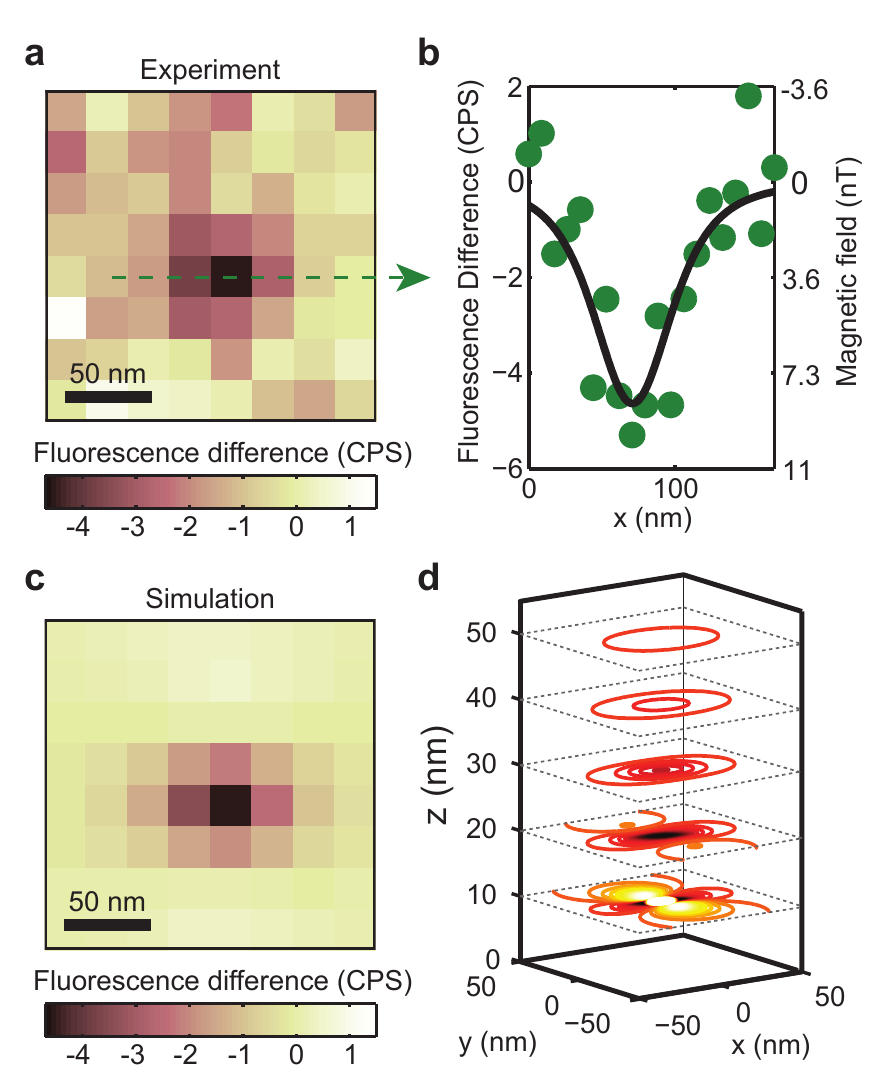}
\caption{\label{Fig4} Single-spin magnetic imaging. (a) Magnetic field image of a target NV spin near the surface of a diamond mesa, acquired with the scanning NV magnetometer. While repeatedly running an AC magnetometry pulse sequence (here with a $32$-pulse XY$8$ sequence, with $40~\mu$s of total evolution time), the sensor NV is laterally scanned over the target, and the fluorescence rates for the target spin starting in the $\ket{0}$ state as well as the $\ket{-1}$ state are independently recorded. Plotted is the difference between these measurements, which depends only on the sensor NV's magnetic interaction with the target spin and not on background fluorescence variations (see Supplementary Information). The pronounced drop in fluorescence near the center of the image indicates a detected single electron spin. (b). An independent magnetometry linecut taken along the green arrow confirms the single spin imaging, which has an intensity and width consistent with the recorded image. The measured fluorescence difference is converted to the measured magnetic field using the sensor NV�s calibrated field sensitivity and fluorescent rate (see Supplementary Fig.\,\ref{FigS5}). (c) Simulated fluorescence due to a target spin. With only the sensor-target displacement as a free parameter, the spin signal is simulated, which agrees well with both the spin image and the linecut for a vertical distance of $51~$nm (the fit in (b) and the image in (c) have the same parameters). (d) If the sensor-target vertical distance can be moderately reduced, the quality of single-spin imaging will be dramatically improved. Plotted are simulated lateral magnetic field contours from a single target electron spin for different sensor-target vertical separations where each contour indicates an increase of signal-to-noise by one for a 100 second integration time. At $50~$nm (the current condition), there is only one contour, indicating single-spin imaging with a signal-to-noise of one; however at $10~$nm, a signal-to-noise of roughly 100 is possible, such that many contours and the dipole lobes of the target spin are clearly observable.}
\end{figure}

Near the center of the magnetometry scan, we observe a drop in the normalized fluorescence from the magnetometry sequence that is well beyond the uncertainty set by the measurement's noise level and is consistent with the effect of a single target NV spin�s magnetic field on the sensor NV. The complete magnetic field image clearly indicates the presence and location of the target NV spin (Fig.\,\ref{Fig4}a). This single electron spin detection is confirmed by repeating the measurement with a spatial linecut of magnetometry measurements (Fig.\,\ref{Fig4}b), with a resulting magnetic response that fits well to a vertical separation of $51.1\pm2.0~$nm between the sensor and target NV centers. (Errors are determined from the $\chi^2$ of the fit as a function of distance, where the sensor-to-target displacement is the only free-parameter and the orientations of both NV spins are independently measured using ESR.) The measured fluorescence difference is converted to a magnetic field at the sensor NV (peak value of $8.6~$nT, Fig.\,\ref{Fig4}b) by using the sensor NV spin's independently calibrated magnetic field response and fluorescence rate. Both scanning magnetometry measurements are in good agreement with simulations of the sensor NV's response to the magnetic field from a single electron spin at a vertical distance of $51~$nm (Fig.\,\ref{Fig4}c). Thus the above measurements are consistent and confirm the detection and nanoscale imaging of the single target spin.

In the demonstrated magnetic field imaging, single-spin measurements with a signal-to-noise ratio (SNR) of one can be acquired in $2.3~$minutes. The data for both single-spin measurements presented in Fig.\,\ref{Fig4} have been integrated for a total time of $42~$minutes per point, yielding an SNR of $4.3$. This integration time is consistent with the measured target NV spin magnetic field ($8.6~$nT) and sensor NV magnetic field sensitivity calculated by assuming the noise is dominated by photon shot noise ($96~$nT/$\sqrt{\rm Hz}$). 
	
By successfully measuring the magnetic field from a single target NV spin, our spin-sensing protocols have been confirmed, enabling otherwise undetectable "dark" electron spins to be detected with confidence. Note that imaging dark spins rather than an NV target could potentially be performed with higher sensitivity, because optical fluorescence from the sensor NV could be better collected and isolated. In addition, for dark spins on or near a sample surface rather than embedded beneath it, the sensor-to-target separation would be reduced by a factor of about two: thus the SNR for magnetic field imaging would increase by nearly an order of magnitude (Fig.\,\ref{Fig4}d) because dipolar fields decay as $\sfrac{1}{r^3}$. Moreover, the required measurement time for a given SNR scales with the sixth power of sensor-to-target separation for a shot-noise-limited measurement: e.g. at a separation of $25~$nm, a target surface spin would be detectable in two seconds (with our instrument's demonstrated sensitivity and a SNR of one).
	
For target spins of interest that cannot be initialized, the variance of the magnetic field at the sensor NV could instead be measured, which is detectable with nearly the same sensitivity as the field itself if an appreciable amount of phase can be acquired\,\cite{Taylor2008}. For instance, at a $25~$nm sensor-to-target distance, with a phase evolution time of $100\mu$s, an uninitialized, driven spin could be detected within two seconds of integration time (SNR of one, for the same sensor NV spin-dependent fluorescence rate and contrast as in the demonstrated spin imaging; Supplementary Fig.\,\ref{FigS6}). For the phase-evolution time used in the demonstrated single-spin imaging ($40\mu$s), the integration time would be $20~$s.
	
If the coherent sensor-target coupling is strong enough for more than $2\pi$ of sensor NV phase to be accumulated during magnetic field measurements, then phase-estimation techniques can be employed, thus allowing the measurement noise to decrease linearly in time\,\cite{Waldherr2012,Nusran2012}, and potentially offering a great boost in speed to magnetic imaging. Moreover, if a target spin can be initialized and has a coherence time as long as the sensor NV, then the target and sensor spins could be entangled. Combined with long-lived storage techniques for quantum states\,\cite{Maurer2012}, the ability to entangle a scanning sensor and target spins could allow for mechanical transfer of quantum information between solid-state spins.
	
Finally, we expect that it will be possible to apply the scanning NV magnetometer to image individual spin targets with high spatial resolution using magnetic field gradients\,\cite{Balasubramanian2008,Grinolds2011}, which can create a nm-scale spatial region where target spins are on electron spin resonance with an applied MW field. By sweeping the frequency of the MW field, individual target spins in different regions could be manipulated and thus detected and mapped spatially by the scanned sensor NV. The spatial resolution of this single-spin magnetic resonance imaging would not be limited by the sensor-to-target separation, and could potentially be pushed to the sub-nm scale with experimentally achieved field-gradients\,\cite{Mamin2007}. Realizing such nano- or atomic scale resolution in imaging single electron spins under ambient conditions would enable diverse applications such as imaging magnetic point defects in solid-state systems\,\cite{Nair2012} and tracking individual spin-labels in biological systems\,\cite{Ayscough1985}.

%\section*{Competing Interests}
%The authors declare that they have no competing financial interests.

%\section*{Correspondence}
Correspondence and requests for materials should be addressed to A.Y. (\href{mailto:yacoby@physics.harvard.edu}{yacoby@physics.harvard.edu})
\\
\\
\\

% Create the reference section using BibTeX: Bibliography appended in the end
\bibliographystyle{apsrev4-1}
\bibliography{manuscript}

\newpage

\newpage

%\begin{widetext}
%\end{widetext}

\onecolumngrid

\section*{Supplementary information:}

\subsection*{Dynamic decoupling of the sensor NV spin:}
	The sensor NV spin coherence time is prolonged by dynamically decoupling it from its noisy environment\,\cite{DeLange2011,DeLange2010,Bluhm2011}. This is achieved by the repeated application of microwave (MW) $\pi$-pulses, which causes the effects of slowly fluctuating magnetic fields to re-phase and cancel out. To apply a large number of pulses without scrambling the sensor NV spin state, the control pulses are carefully calibrated to within $2\%$ using a boot-strap tomography scheme\,\cite{Dobrovitski2010}. For the dynamic decoupling scheme and magnetometry, we employ an XY8 sequence\,\cite{Gullion1991}, which uses �-pulses around two orthogonal axes on the equator of the sensor NV�s Bloch sphere to minimize the accumulation of pulse errors. This sequence ($\pi_x-\pi_y-\pi_x-\pi_y-\pi_y-\pi_x-\pi_y-\pi_x$) is repeated as many times as possible to maximize the sensor NV's magnetometry sensitivity, which -- as a function of the number of pulses -- is a compromise between the extended NV coherence from the decoupling and the reduced contrast from accumulated pulse errors. 

	MW fields are supplied from a Rhode and Schwarz SMB100A signal generator. MW phase control is achieved using an IQ mixer (Marki-1545) with pulsed analog inputs on the I and Q ports supplied by an arbitrary waveform generator (Tektronix AWG5000). NV spin Rabi frequencies in this work are $15$-$20~$MHz, with typical $\pi$-pulse durations of $30~$ns.

\subsection*{Adiabatic fast passages for controlling the target NV spin:}

	To control the target NV spin with high fidelity over numerous spin inversions, we employ adiabatic fast passages. The spin-state is prepared optically in the mS = |0> state, and microwaves (MW) with bare Rabi frequency $\omega_R$ are applied and detuned by $\delta(t=0)$ from the target NV transition. The detuning is ramped through zero to $-\delta(t=0)$ over a pulse time, $T_p$. At any point in time, the target NV spin in the rotating frame prececess around an effective magnetic field $\Omega_R$, which is the sum of the MW field and the remaining static magnetic field in the rotating frame resulting from the non-zero MW detuning. If the angular velocity, d$\theta/$dt of $\Omega_R$ is slow compared to $\omega_R$, then the NV spin-state is effectively locked to the motion of this effective magnetic field as it moves from $\ket{0}$ to $\ket{-1}$. In general, it is advantageous to sweep the detuning non-linearly in time and spend most of TP when the NV spin is near the equal population state where it is most susceptible to dephasing\,\cite{Garwood2001}. To achieve this, we ramp the detuning to keep the rate of change of the spin's angle with respect to the $\,\ket{0}$ state constant, so that:
\begin{equation}
\delta(t)=\omega_R \tan(\beta(\frac{t}{2T_p}-1))
\end{equation}	
where $\beta$ is chosen to achieve the desired sweep range. For the adiabatic fast passages presented in Fig.\,\ref{Fig3} of the main text, $T_p=300~$ns, $\delta(t=0)=100~$MHz, and $\omega_R=17~MHz$. 

The detuning ramping is implemented by using an arbitrary waveform generator to output a sinusoid at a frequency of the desired detuning, which is mixed with a continuous-wave MW source (all are the same make and model as the sensor-addressing MW equipment). By setting the phase of this sinusoid to be the integral of the detuning as a function of time, the mixed MW frequency can be continuously varied.

\setcounter{figure}{0}

\makeatletter
\renewcommand{\thefigure}{S\@arabic\c@figure}

\begin{figure*}
\includegraphics[scale=1]{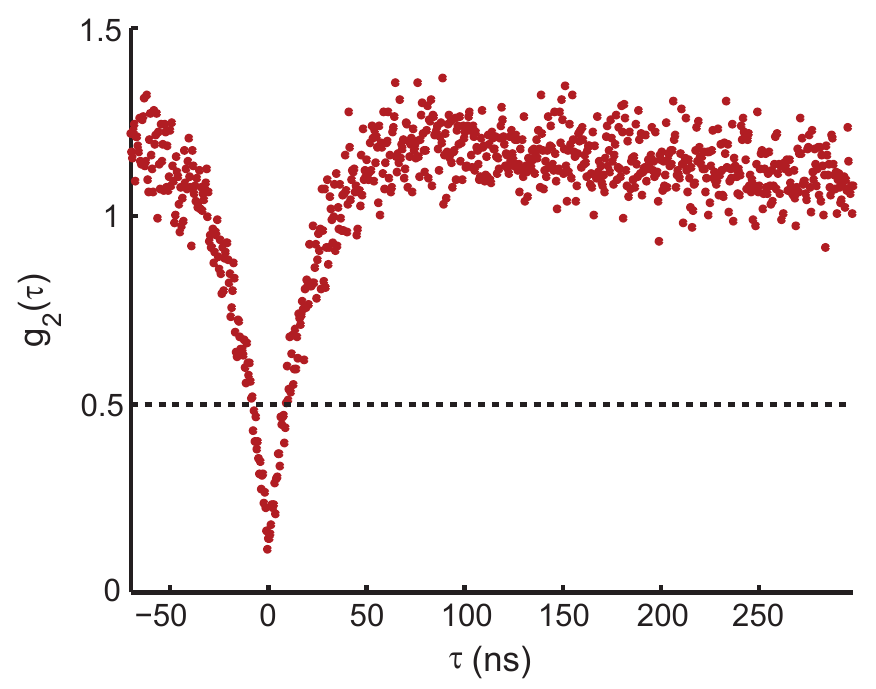}
\caption{\label{FigS1} Photon-autocorrelation measurements for the target NV center: Photon-autocorrelation measurements for the target NV (in the absence of the sensor NV) give $g^2(\tau=0)<0.5$, ensuring that a single target NV spin lies in the diamond mesa on which magnetic field imaging is performed with the scanning NV magnetometer. No background subtraction was performed on the data and normalization was performed based on count-rates on the individual detectors as well as the time-binning in the photon-correlation hardware.}
\end{figure*}

\begin{figure*}
\includegraphics[scale=1]{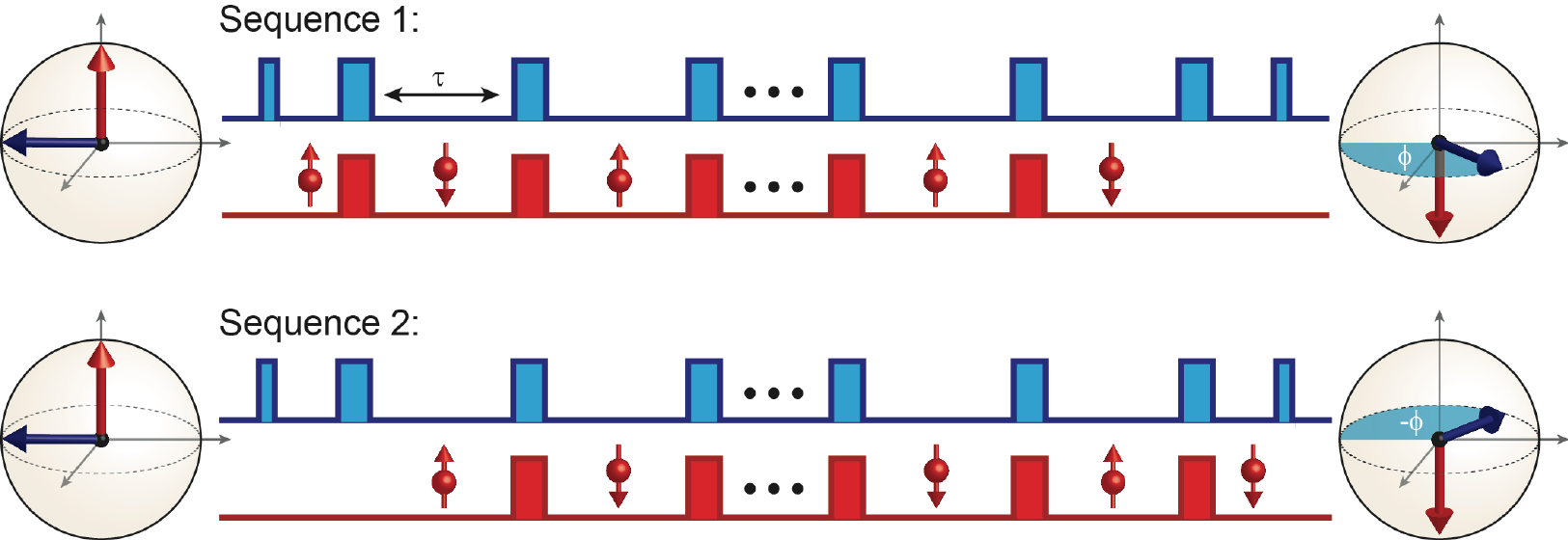}
\caption{\label{FigS2} Magnetometry normalization scheme: To isolate the sensor NV�s spin-state-dependent fluorescence from spatially varying background fluorescence, we employ a normalization scheme that involves alternatively performing two slightly different magnetometry pulse sequences, and then subtracting the measured NV fluorescence rates. These two sequences are similar to the double-electron-electron resonance scheme presented in Fig.\,\ref{FigS3}, except that one target NV spin-inversion (performed via an adiabatic fast passage) is removed from each sequence: in Sequence $1$, the last spin inversion is removed, while in Sequence 2 it is the first spin inversion. By removing spin inversions in this manner, we ensure that the sensor NV (blue arrow) acquires an equal and opposite target-spin-induced phase shift ($\Phi$, shaded blue region) during the two pulse sequences, because the target spin-state is inverted in the two sequences for the majority of the phase evolution time. Crucially, the target NV spin ends in the same state for the two pulse sequences (here, $\ket{-1}$). Thus subtracting the measured NV fluorescence rates for the two pulse sequences removes the contribution of background fluorescence from the NV target, which has a non-trivial spatial dependence (Fig.\,\ref{Fig2} of the main text). Moreover, both pulse sequences have the same number of target spin inversions, which alleviates unavoidable spin-polarization losses associated with flipping the target spin and cross-talk between the applied MW sources. In this scheme, a small amount of integration time is unused for sensor NV phase accumulation (the portions next to the $\sfrac{\pi}{2}$ pulses at the beginning and end of each sequence cancel, equal to one delay period $\tau$ between $\pi$ pulses). However in the limit of a large number of spin inversions, this loss of integration time is negligible.
}
\end{figure*}

\begin{figure*}
\includegraphics[scale=1]{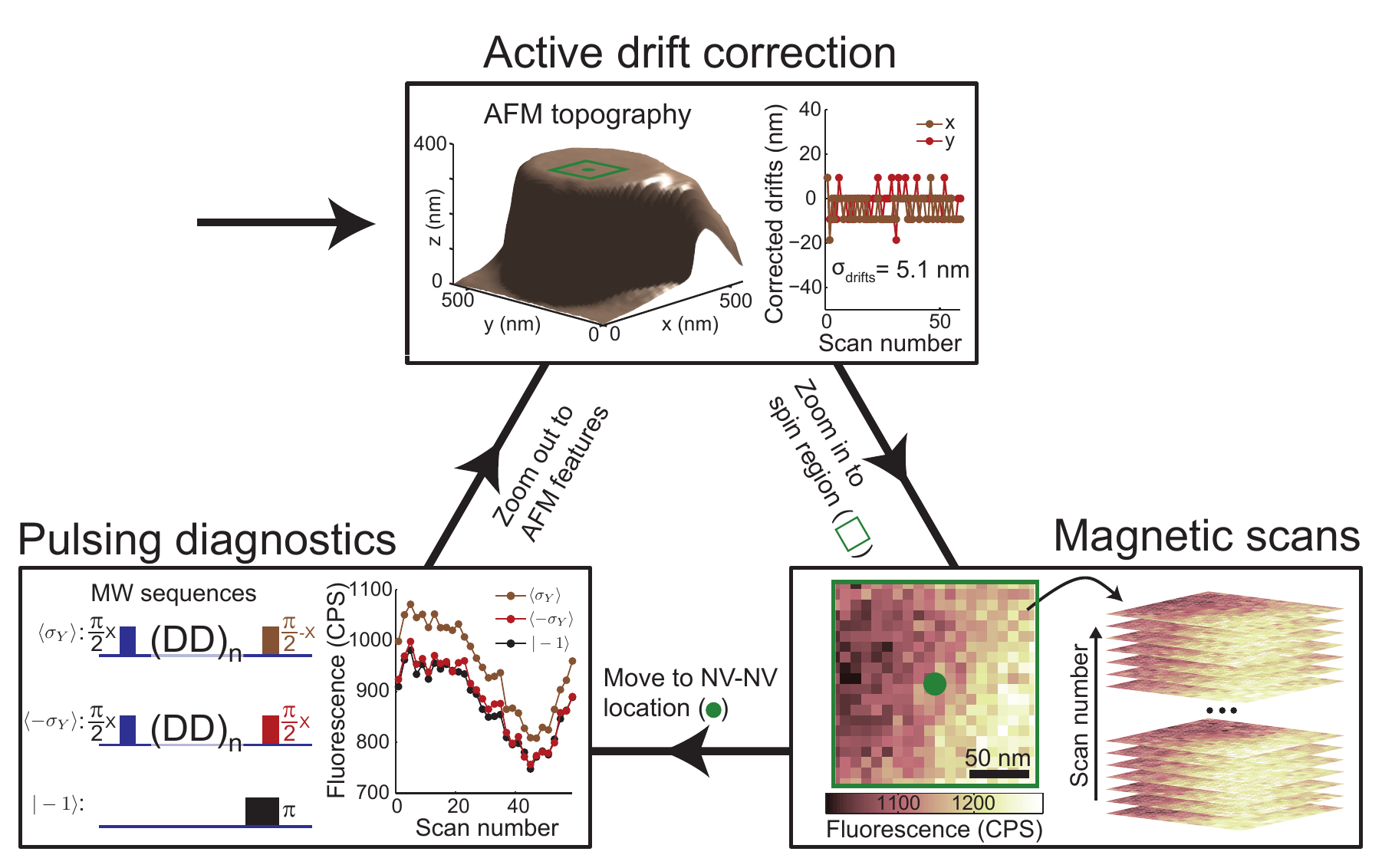}
\caption{\label{FigS3} Magnetic field image acquisition protocol: Scanning an AFM for long periods of time with nanometer precision can be difficult to achieve under ambient conditions because of thermal-induced drifts. Temperature fluctuations on the order of a fraction of a degree can lead to tens of nanometers of relative motion between the sensor and target NV spins, which would considerably smear out our magnetic field imaging. These drifts generally occur on long time-scales, with a few nanometers of drift every hour. To minimize their effect, we employ an image acquisition protocol that periodically corrects for sensor-to-target drifts.\\
The protocol has three major components: (1) Drift-correction using the sample topography to determine the target's location (top panel); (2) Taking a relatively quick magnetic scan over the target spin location (lower right panel); and (3) Checking to make sure that the MW pulsing has not degraded over time, and that the sensor NV�s magnetic sensitivity has not been significantly compromised (lower left panel).\\
Drift correction is performed by scanning over the target-containing diamond mesa, and the measured topography is the convolution of this mesa ($\sim200~$nm in diameter) with the diamond nanopillar scanning tip ($\sim200~$nm  in diameter). From this topography, and the simultaneously measured fluorescence (as in Fig.\,\ref{Fig2} of the main text), the target NV spin can be located (green dot) and an appropriate scan range can be defined (green square). The topography of successive scans (taken after both magnetometry and diagnostic measurements), can be compared to the first reference scan by cross-correlation, and thus drifts can be corrected between scans. In the single-spin measurements presented in Fig.\,\ref{Fig4} of the main text, we observe a mean variation of $5~$nm between successive scans (limited by the pixel size of the reference scan), indicating that magnetic field images can be overlapped with roughly $5$-nm precision.\\
After zooming into the appropriate scan region, where the expected target spin NV lies in the center of the scan range, magnetic field images are acquired while simultaneously alternating between the two magnetic detection pulse sequences (Fig.\,\ref{FigS2}) and monitoring their fluorescence rates (only one sequence is illustrated). Each scan is integrated for roughly $30~$minutes to minimize the drifts between scans.\\
When a magnetic scan is finished, the sensor NV is placed at the approximate measurement position to measure the optimal sensitivity to the target NV and as a function of time. In general, the sensor NV can slowly drift in and out of the green laser confocal spot, causing variations in the overall detected NV fluorescence. Additionally, the power of the MW source can drift, which can decrease the performance of dynamic decoupling and magnetometry pulse sequences. Magnetic field sensitivity is experimentally determined by running the magnetometry sequence, with the phase of the last $\sfrac{\pi}{2}$ pulse set at $\pm\sfrac{\pi}{2}$ (red and brown data points, respectively) to measure the $\left<-\sigma_y\right>$ and $\left<\sigma_y\right>$ projections of the sensor NV. The difference between these measurements gives the contrast and counts of the sensor NV�s magnetic response, and when combined with the phase accumulation time, determines the magnetic field sensitivity of the sensor NV. To differentiate overall NV fluorescence rate changes from pulsing performance changes, we also measure fluorescence counts for the $\ket{-1}$ state (black data points), which should overlap with the $-\sigma_y$ measurement in the case of no pulse errors or dephasing.\\
After these pulse diagnostics, we zoom out to measure the topography of the sample again, completing a measurement cycle. This procedure is repeated until a desired signal-to-noise in the magnetic field image has been achieved. 
}
\end{figure*}

\begin{figure*}
\includegraphics[scale=1]{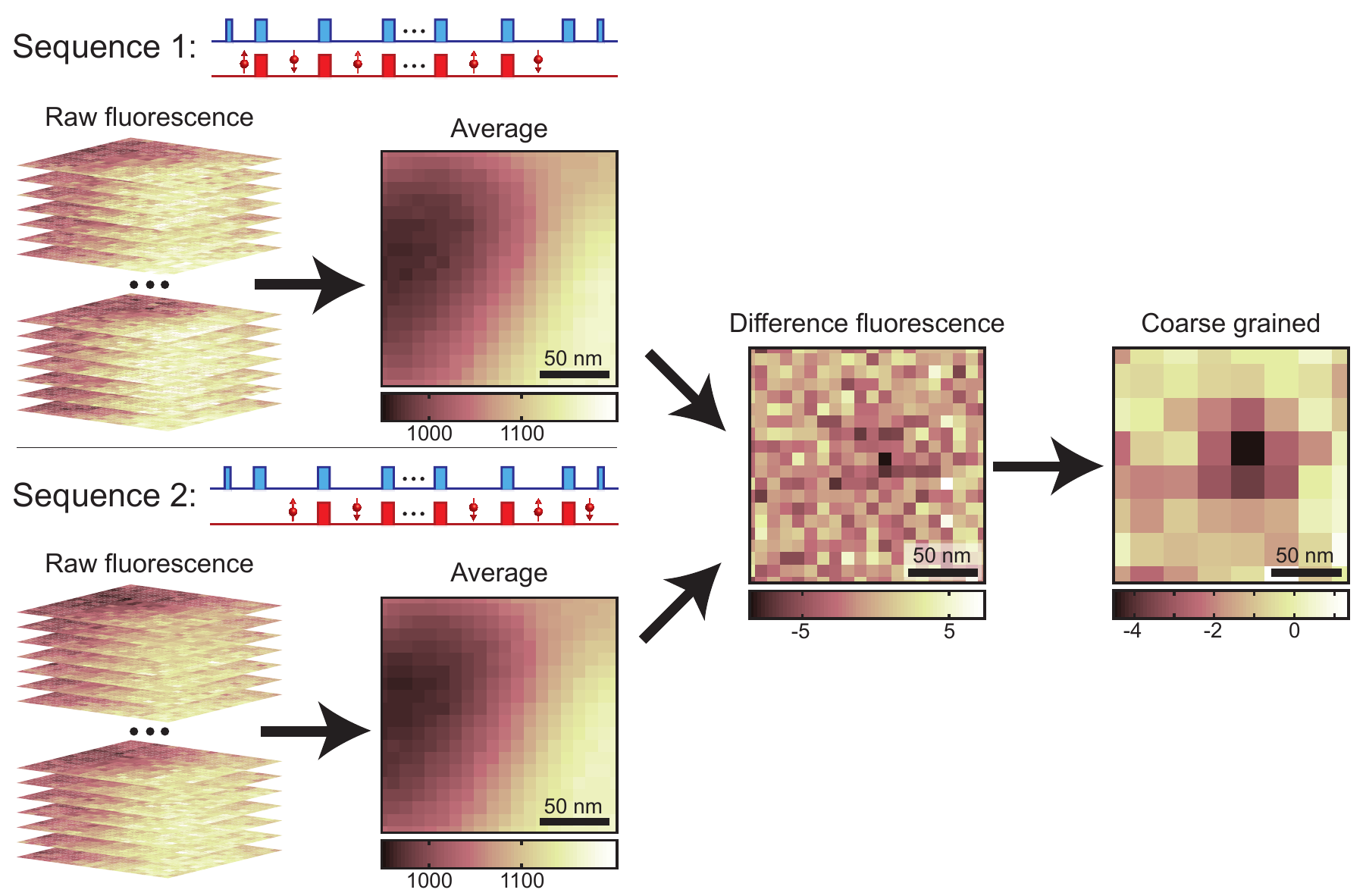}
\caption{\label{FigS4} Magnetic field image data processing: The multiple images taken during a magnetic field scan for each magnetometry sequence (left column) are averaged (without any further spatial correction), to yield the average fluorescence map (center left column). In these averaged measurements, there are large variations in fluorescence due to near field coupling into and out of the diamond nanopillar for the target and sensor NV centers, as described in Fig.\,\ref{Fig2} of the main text. These variations are quite large ($\sim150~$CPS) compared to the expected effect of a single target NV�s magnetic field on the sensor NV signal (fluorescence change $\sim4~$CPS under inversion of the target NV spin). Subtracting the average fluorescence maps of the two magnetometry sequences yields a difference fluorescence signal free of the large background fluorescence (center right). In general, the difference in fluorescence between the two magnetometry sequences has a small remaining offset, and so it has a mean of a few counts per second, even in the absence of the target NV. This is likely due to a small amount of cross-talk between the target-addressing MW and the sensor NV, which is slightly different between the two sequences. During our pulsing diagnostics in the measurement acquisition, we measure this fluorescence offset with the sensor NV very far from the sample ($>1\mu$m), and we subtract this value from the difference fluorescence, which yields a mean of zero counts per second away from the target NV spin. As long as this remainder fluorescence ($4.3~$CPS for this image) times the percent fluorescence variations across the scan region ($15\%$) is smaller than the target NV spin signal � as is the case here � then the target NV spin signal will be the largest feature in the difference fluorescence map. To increase the signal-to-noise ratio (SNR) of magnetic field imaging, we average multiple pixels together to coarse-grain the scan (right panel), yielding a scan with $64$ pixels across a field-of-view of $\sim200\times200~$nm, with $42~$minutes of integration time per pixel providing average SNR of $4.3$.
}
\end{figure*}

\begin{figure*}
\includegraphics[scale=1]{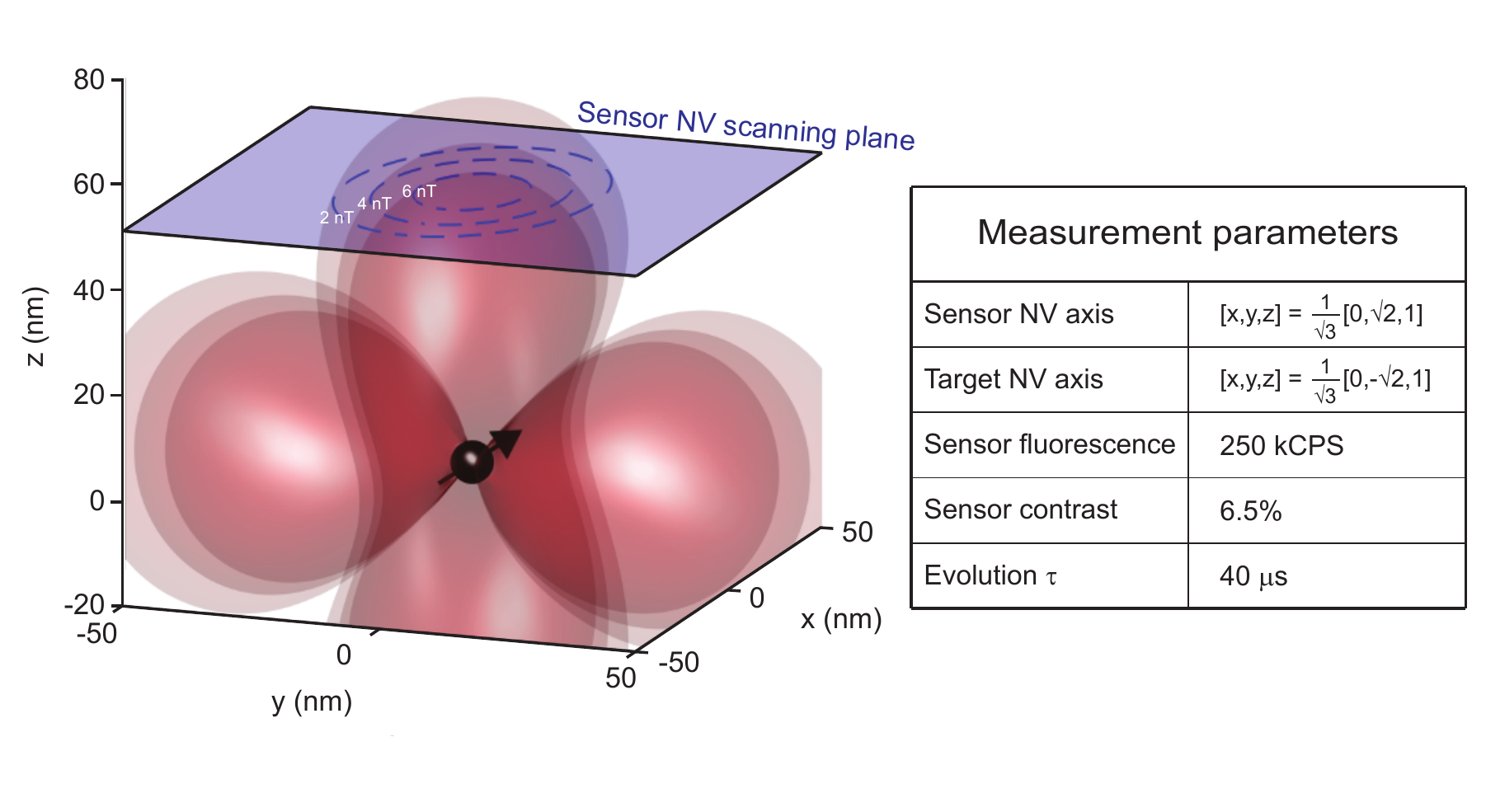}
\caption{\label{FigS5} Simulation of single-spin magnetic field imaging: The response of the scanning NV magnetometer to a single electronic spin is simulated by considering an electron spin at the origin of a coordinate system with a quantization axis oriented along the $[x,y,z] = [0,-\sqrt{2},1]$ direction (to match the orientation in the sample that is measured through ESR measurements using a three-axis Helmholtz coils). Equi-field contours of this target spin�s magnetic field are plotted as a function of three-dimensional space (displayed here are $2~$nT, $4~$nT, and $6~$nT). Because the sensor NV is first-order sensitive only to magnetic fields along its quantization axis ($[x,y,z] = [0,\sqrt{2},1]$), the plotted field contours from the target spin have been projected along the sensor NV quantization axis, which yields the dipole field lobe pattern shown here. Experimental magnetic field scans are taken as plane-cuts of this dipole field pattern above the location of the target spin (at an a-priori unknown distance; plotted is the best fit value of $51~$nm). The simulated magnetic field profile is converted into a spatial map of sensor NV fluorescence rate using measured values of the sensor NV fluorescence rate, contrast, and phase evolution time, giving a magnetic field sensitivity conversion factor of $-1.8~$nT per count per second.
}
\end{figure*}

\begin{figure*}
\includegraphics[scale=1]{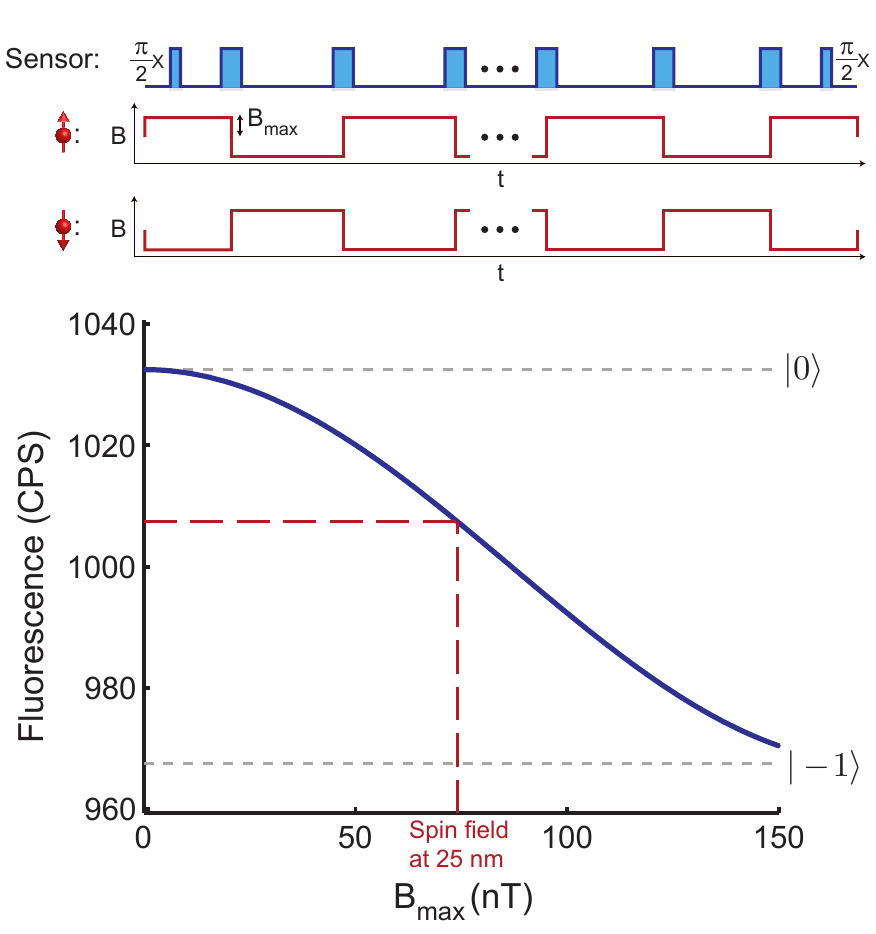}
\caption{\label{FigS6} Measuring the variance of a non-initialized spin: If target electron spins cannot be initialized (unlike the target NV spin measured in this work), then the spin's magnetic field will average out to zero over multiple measurements, as at the start of a given measurement (�shot�) the target spin has an equal probability of being either up or down. However, if the target's spin�s longitudinal relaxation time is much longer than the magnetometry phase-evolution time $\tau$, within a single shot, then the target spin maintains its statistical polarization, and a net phase shift will be accumulated by the sensor spin. By choosing the axis of rotation of the final $\sfrac{\pi}{2}$ pulse to match the axis of the first $\sfrac{\pi}{2}$ pulse, when the accumulated phase shifts from multiple measurement shots are converted to a net population difference, the effect of the target spin�s magnetic field no longer cancels out and can be measured via the sensor NV's spin-dependent fluorescence. This scheme effectively measures the variance of the target spin polarization ($\left<\sigma_z^2\right>-\left<\sigma_z\right>^2$, for the thermal state of a target spin) instead of its mean polarization ($\left<\sigma_z\right>$)
	Plotted here is the sensor NV�s response for $\tau=100\mu$s, plotted is the sensor's NV response to a (driven) target electron spin with random polarization (either up or down) at the measurement's start. (The sensor NV�s fluorescence and spin-dependent contrast used are those demonstrated in spin-imaging; Supplementary Fig.\,\ref{FigS5}.). The magnetic field profile for this driven target spin is a square wave with amplitude $B_{\rm max}$, which is synchronized to the sensor NV's decoupling scheme. For a sensor-to-target distance of $25~$nm (and the same sensor and target spin-quantization axes used in the present work; Supplementary Fig. 5), $B_{\rm max}=74~$nT, which gives a signal of $25~$CPS with respect to the $\ket{0}$ state. Within two seconds of integration time, this signal divided by the measurement's shot noise gives a signal to noise ratio of one.
}
\end{figure*}

\end{document}